\documentclass[aps,twocolumn,superscriptaddress,floatfix]{revtex4}
\usepackage{psfig}
\begin{document}

\newcommand{\bed}{\[}
\newcommand{\eed}{\]}
\newcommand{\beq}{\begin{equation}}
\newcommand{\eeq}{\end{equation}}
\newcommand{\beqa}{\begin{eqnarray}}
\newcommand{\eeqa}{\end{eqnarray}}
\newcommand{\ket} [1] {\vert #1 \rangle}
\newcommand{\bra} [1] {\langle #1 \vert} 
\newcommand{\braket}[2]{\langle #1 | #2 \rangle}
\newcommand{\proj}[1]{\ket{#1}\!\!\bra{#1}}
\newcommand{\mean}[1]{\langle #1 \rangle}

\title{Quantum cloning of orthogonal qubits}

\author{J. Fiur\'{a}\v{s}ek}
\affiliation{Department of Optics, Palack\'{y} University, 17. listopadu 50,
77200 Olomouc, Czech Republic}
\author{S. Iblisdir}
\affiliation{Ecole Polytechnique,
CP 165, Universit\'{e} Libre de Bruxelles, B-1050 Bruxelles, Belgium}
\author{S. Massar}
\affiliation{Service de Physique  Th\'{e}orique, CP 225,
Universit\'{e} Libre de Bruxelles, B-1050 Bruxelles, Belgium}
\author{N.J. Cerf}
\affiliation{Ecole Polytechnique,
CP 165, Universit\'{e} Libre de Bruxelles, B-1050 Bruxelles, Belgium}

\begin{abstract}
An optimal universal cloning transformation is derived that produces 
$M$ copies of an unknown qubit from a pair of orthogonal qubits.
For $M>6$, the corresponding cloning fidelity is higher than that
of the optimal copying of a pair of identical qubits. It is shown
that this cloning transformation can be implemented probabilistically
via parametric down-conversion by feeding the signal and idler
modes of a nonlinear crystal with  orthogonally polarized photons.
\end{abstract}

\maketitle

In contrast to classical information, quantum information cannot be
copied. This so-called {\em no-cloning} theorem \cite{Wootters82}, which
is a direct consequence of the linearity of quantum theory,
states that it is impossible to prepare several exact copies (or clones)
of an unknown quantum state $|\psi\rangle$. 
Although exact cloning is forbidden, one can design various 
quantum cloning machines which produce approximate clones.
In particular, much attention has been devoted to
the optimal universal cloning machines for qubits,
which prepare $M$ identical approximate clones out of $N$ replicas
of an unknown qubit, and such that the fidelity of the clones 
is state-independent \cite{Buzek96}. 
Cloning machines for states in a $d$-dimensional Hilbert
space (qudits) were also investigated \cite{Buzek98},
as well as continuous-variable cloning machines
for coherent states \cite{Cerf00}.

In the limit of an infinite number of clones, the optimal cloning reduces to
the optimal quantum measurement. In this context, a very interesting
observation has been made by Gisin and Popescu \cite{Gisin99}
who noted that the information about a direction in space is better encoded 
into two orthogonal qubits than in two identical ones. If we possess
a two-qubit state $|\psi,\psi_\perp\rangle$ with 
$\langle\psi|\psi_\perp\rangle=0$, then we can estimate $|\psi\rangle$ 
with a fidelity $F_\perp= (1+1/\sqrt{3})/2 \approx 0.789$ 
\cite{Gisin99,Massar00}. This slightly exceeds the fidelity of
the optimal measurement on a qubit pair $|\psi,\psi\rangle$, $F_{||}=3/4$.
A similar situation occurs for continuous quantum
variables. Suppose we want to encode a (randomly chosen)
position in phase space. A possible strategy would be to prepare
a pair of coherent states $|\alpha,\alpha\rangle$, where the real
and imaginary parts of the complex number $\alpha$ represent
the phase-space coordinates. However, it is actually better 
to supply a state $|\alpha,\alpha^\ast\rangle$ from which $\alpha$ can
be inferred  via optimal measurement with a lower error \cite{Cerf01a}.
It can also be shown that the state $|\alpha,\alpha^\ast\rangle$ gives
an advantage when cloning coherent states: $M$ identical approximate clones
of a coherent state $|\alpha\rangle$ can be prepared with a higher fidelity
from the state $|\alpha,\alpha^\ast\rangle$ than from
$|\alpha,\alpha\rangle$ \cite{Cerf01b}.

Motivated by this result, we were led to ask whether a similar
situation might also occur for qubits.
Can $M$ clones of qubit $|\psi\rangle$ be produced from
an orthogonal qubit pair $|\psi,\psi_\perp\rangle$
with a higher fidelity than from an identical pair $|\psi,\psi\rangle$?
In this Letter, we answer this question by an affirmative.
We present a universal cloning machine acting on an orthogonal qubit pair
that approximately implements the transformation
$|\psi\rangle|\psi_\perp\rangle \rightarrow 
|\psi\rangle^{\otimes M}$ with the optimal fidelity.
Then, we show that this cloning transformation 
can be implemented probabilistically in quantum optics
by use of parametric down-conversion. Our proposed setup 
extends the scheme of Simon {\em et al.} \cite{Simon00}
by feeding both the signal and idler modes of a nonlinear crystal
with $|\psi\rangle$ and $|\psi_\perp\rangle$, respectively.

Let us first provide a simple argument on why we can expect the state
$|\psi,\psi_\perp\rangle$ to be better cloned than $|\psi,\psi\rangle$.
If we perform an optimal measurement of $|\psi,\psi_\perp\rangle$, we can 
prepare $M$ identical clones of $|\psi\rangle$, each with a
fidelity $F_{\perp}$. In contrast,
the fidelity of the optimal universal cloning machine that prepares 
$M$ clones from a two-qubit state $|\psi,\psi\rangle$
is given by $F_{||}(M)=(3M+2)/(4M)$ \cite{Buzek96}. 
Clearly, $F_{||}(M)<F_\perp$ for sufficiently large $M$.
Hence, this (non-optimal) measurement-based cloning of 
$|\psi,\psi_\perp\rangle$ is better than the standard
cloning of $|\psi,\psi\rangle$.

Let us now seek for a unitary transformation which 
{\em optimally} 
approximates the transformation $|\psi\rangle|\psi_\perp\rangle 
\rightarrow |\psi\rangle^{\otimes M}$. Since the set of all states 
of the form $\ket{\psi} \ket{\psi_{\perp}}$ span the whole Hilbert space of 
two qubits, the most general transformation is of the form:
\beq\label{eq:gentsf}
\ket{i}\ket{j}\ket{R} \to \sum_{k=0}^{M} \ket{M,k}\ket{R_{ijk}} \qquad i,j=0,1
\eeq
where $\ket{R}$ and $\ket{R_{ijk}}$ respectively denote the initial and 
final states of the ancilla, while $|M,k\rangle$ ($k=0,\ldots,M$)
denotes a symmetric $M$-qubit state with $k$ qubits in state $|0\rangle$ 
and $M-k$ qubits in state $|1\rangle$. 
The arbitrary state of a qubit $\ket{\psi}$ can be conveniently 
written as $\ket{\psi}=d(\Omega)\ket{0}=\sum_i d_{i0}(\Omega)\ket{i}$, 
where the matrix $d(\Omega)$ is given by
\beq
d(\Omega)=\left( \begin{array}{cc} \cos\frac{\vartheta}{2}
 & e^{-i\phi} \sin\frac{\vartheta}{2} \\[1mm]
e^{i\phi}\sin \frac{\vartheta}{2} & -\cos\frac{\vartheta}{2}
\end{array} \right),
\eeq 
with $\vartheta$ and $\phi$ denoting the usual polar and azimuthal
angles pointing in direction $\Omega$.
The linearity of (\ref{eq:gentsf}) implies that 
an arbitrary pair of orthogonal qubits transforms according to
\beq
\ket{\psi}\ket{\psi_{\perp}} \to \ket{\Psi_{\rm out}(\psi)}=\sum_{ijk} 
d_{i0}(\Omega) d_{j1}(\Omega) \ket{M,k}\ket{R_{ijk}}.
\eeq
We will measure the quality of the transformation by the average single-clone 
fidelity $F_{\perp}(M)$. Denoting by ${\rm Tr}_{1,\rm anc}$ the partial trace 
over the ancilla and all the clones but the first one, we get
\begin{eqnarray}
F_{\perp}(M) &=&\int d\Omega \bra{\psi}\,{\rm Tr}_{1,{\rm anc}} 
[\ket{\Psi_{\rm out}(\psi)}
\bra{\Psi_{\rm out}(\psi)}]\, \ket{\psi} 
\nonumber \\
&=&
\sum_{i'j'k'} \sum_{ijk}\braket{R_{i'j'k'}}{R_{ijk}} A^{i'j'k'}_{ijk},
\label{fidelityform}
\end{eqnarray}
where 
\beqa
&&A^{i'j'k'}_{ijk}=\sum_{n,n'} 
\bra{n'} {\rm Tr}_1 [\,\ket{M,k}\bra{M,k'}\,]\ket{n}
\nonumber\\
&&\times \int d\Omega \,  d_{n0}(\Omega) d^{*}_{n'0}(\Omega) 
d_{i0}(\Omega) d_{j1}(\Omega) d^{*}_{i'0}(\Omega) d^{*}_{j' 1}(\Omega) . 
\nonumber\\
\eeqa
The coefficients $A_{ijk}^{i'j'k'}$ can be considered as matrix 
elements of an operator $A$ acting on the space ${\cal{H}}\otimes\cal{K}$,
where $\cal{H}$ denotes the Hilbert space of the two input qubits
and $\cal{K}$ denotes the Hilbert space of symmetric states of $M$ output
qubits.  Similarly, 
$\chi_{ijk}^{i'j'k'}=\langle R_{ijk}|R_{i'j'k'}\rangle$ define matrix elements
of an operator
$\chi$ also acting on ${\cal{H}}\otimes\cal{K}$. The formula 
(\ref{fidelityform}) for the fidelity thus simplifies to $F_\perp(M)={\rm
Tr}_{{\cal{H}},\cal{K}}[\chi A]$. The operator $\chi$ uniquely represents the
completely positive cloning map, which transforms operators supported on 
$\cal{H}$ onto operators supported on $\cal{K}$. By definition, the operators
$A$ and $\chi$ are Hermitian and positive semidefinite, 
$A\geq 0$ and $\chi\geq 0$.

Of course, the transformation (\ref{eq:gentsf}) should be unitary, which reads
$\sum_{k} \braket{R_{i'j'k}}{R_{ijk}}=\delta_{i'i}\delta_{j'j}$.
This is equivalent to ${\rm Tr}_{\cal{K}}[\chi]= \openone_{\cal{H}}$,
where $\openone_{\cal{H}}$ is the identity operator on $\cal{H}$.
Thus, introducing a set of Lagrange multipliers $\lambda^{i' j'}_{i j}$ 
for these unitarity constraints, our problem amounts to 
extremize the quantity $W={\rm Tr}_{{\cal{H}},\cal{K}}[(A-\Lambda)\chi]$
under the constraint $\chi\geq 0$,
where $\Lambda=\lambda\otimes \openone_{\cal{K}}$ 
and $\lambda$ is the matrix of Lagrange multipliers
($\openone_{\cal{K}}$ is the identity operator on $\cal{K}$). 
Varying $W$ with respect to the eigenstates of the operator $\chi$, 
we get the extremal equation 
\begin{equation}
(A-\Lambda)\chi=0
\end{equation}
for the optimal $\chi$.
Following \cite{Fiurasek01b}, this equation can be further transformed into a
form suitable for numerical calculation via repeated applications of 
\begin{eqnarray}
\chi=\Lambda^{-1}A\chi A\Lambda^{-1}, \qquad
\lambda=({\rm Tr}_{\cal{K}}[A\chi A])^{1/2}.
\label{extremaleq}
\end{eqnarray}
Note that the matrix $\lambda \geq 0$ is determined from the
unitarity constraints. 

By numerically solving Eq. (\ref{extremaleq}) for $M=2,\ldots 15$,
we have been able to conjecture the general analytical form 
of the optimal transformation:
\begin{equation}
|\psi,\psi_\perp\rangle \rightarrow
\sum_{j=0}^M \alpha_{j,M} |(M-j)\psi,j\psi_\perp\rangle
\otimes |(M-j)\psi_\perp,j\psi\rangle,
\label{cloner}
\end{equation}
where 
\begin{equation}
\alpha_{j,M}=(-1)^j \left[\frac{1}{\sqrt{2(M+1)}}
                    +\frac{\sqrt{3}(M-2j)}{\sqrt{2M(M+1)(M+2)}}\right],
\label{alfa}
\end{equation}
with $| j\psi,(M-j)\psi_\perp\rangle$ denoting a totally symmetric state
of $M$ qubits where $j$ qubits are in state $|\psi\rangle$ and $M-j$
qubits are in state $|\psi_\perp\rangle$. The first $M$ output qubits contain
the clones of state $|\psi\rangle$ while the other $M$ qubits contain
the clones of $|\psi_\perp\rangle$ (or anticlones). 

We stress here that the cloning transformation (\ref{cloner}) is unitary.
Since this is by no means obvious from (\ref{cloner}),
let us present a proof of this.
We can expand any state $|j\psi,(M-j)\psi_\perp\rangle$
in the basis $|M,k\rangle$ as
\begin{equation}
|j\psi,(M-j)\psi_\perp\rangle =
\sum_{k=0}^M e^{i(j-k)\phi}D_{kj}^M(\vartheta) |M,k\rangle.
\label{expansion}
\end{equation}
We will not need an explicit expression for the functions
$D_{kj}^M(\vartheta)$ here, but will only use some of their properties.
Since the functions $D_{kj}^M(\vartheta)$ are elements 
of a (real) unitary matrix, they satisfy the orthogonality relation,
\begin{equation}
\sum_{j=0}^M D_{kj}^M(\vartheta)D_{lj}^M(\vartheta)=\delta_{kl}.
\label{orthogonality}
\end{equation}
We will also use the following recurrence formula \cite{Vilenkin91},
\begin{widetext}
\begin{equation}
(2j-M)D_{kj}^M(\vartheta)=(2k-M) \cos\vartheta \, D_{kj}^M(\vartheta)
+\sin \vartheta \sqrt{(k+1)(M-k)}\, D_{k+1,j}^M(\vartheta)
+\sin\vartheta\sqrt{k(M-k+1)}\, D_{k-1,j}^M(\vartheta) .
\label{recurrence}
\end{equation}
\end{widetext}
For the purposes of the proof it is convenient to apply a unitary
transformation $U_0$ on the last $M$ qubits at the output of the cloner
and get the state $|\Phi_{\rm out}(\psi)\rangle=
I^{\otimes M}\otimes U_0^{\otimes M}|\Psi_{\rm out}(\psi)\rangle$.
The unitary transformation $U_0$ flips the states $|0\rangle$ and $|1\rangle$,
$|0\rangle\rightarrow |1\rangle$ and $|1\rangle\rightarrow -|0\rangle$.
Thus
$|(M-j)\psi_\perp,j\psi\rangle \rightarrow
(-1)^j|(M-j)\psi^\ast,j\psi_\perp^\ast\rangle$
where
$|\psi^\ast\rangle=\sum_i d_{i0}^\ast|i\rangle$.
Next we expand $|(M-j)\psi,j\psi_\perp\rangle$ and
$|(M-j)\psi^\ast,j\psi_\perp^\ast\rangle$ in the basis $|M,k\rangle$
according to Eq. (\ref{expansion}), and then
utilize the recurrence formula (\ref{recurrence}). Finally,
we can carry out the sum over $j$ with the help of
Eq. (\ref{orthogonality}), resulting in
\begin{widetext}
\begin{eqnarray}
&&|\Phi_{\rm out}(\psi)\rangle=\sum_{k=0}^M 
\left[ a_M + b_{M}(2k-M)\right] \cos^2\frac{\vartheta}{2}
|M,k\rangle\otimes |M,k\rangle
+\sum_{k=0}^M \left[ a_M - b_{M}(2k-M)\right] \sin^2\frac{\vartheta}{2}
|M,k\rangle\otimes |M,k\rangle
\nonumber \\
&&+e^{i\phi} \sum_{k=0}^M b_M \sin\vartheta \sqrt{(M-k)(k+1)}
|M,k\rangle\otimes |M,k+1\rangle
+e^{-i\phi}\sum_{k=0}^M b_M \sin \vartheta \sqrt{k(M-k+1)}
|M,k\rangle \otimes |M,k-1\rangle,
\nonumber \\
\label{Psifinal}
\end{eqnarray}
\end{widetext}
\vspace*{-12mm}
where the coefficients $a_M$ and $b_M$ read
\[
a_M=\frac{1}{\sqrt{2(M+1)}}, \qquad
b_M=\frac{\sqrt{3}}{\sqrt{2M(M+1)(M+2)}}.
\]
The four terms on the right-hand
side of Eq. (\ref{Psifinal}) are proportional to
the output states for the four input basis states $|01\rangle$, $|10\rangle$,
$|00\rangle$, and $|11\rangle$, respectively.
Consequently, it is easy to prove that
the transformation $|\psi,\psi_\perp\rangle\rightarrow
|\Phi_{\rm out}(\psi)\rangle$
preserves scalar products, hence is unitary.

Let us now calculate the fidelity of the clones. We can see from 
Eq. (\ref{cloner}) that the cloning machine preserves the symmetry
of the input state $|\psi,\psi_\perp\rangle$, so
the clones of both states $|\psi\rangle$ and
$|\psi_\perp\rangle$ have the same fidelity. This state-independent
single-qubit fidelity can be obtained by summing a series,
\begin{equation}
F_{\perp}(M)= \sum_{j=0}^M \frac{M-j}{M} \, \alpha_{j,M}^2.
\label{FortMsum}
\end{equation}
After some algebra, we arrive at the expression
\vspace*{-2mm}
\begin{equation} 
F_{\perp}(M) = \frac{1}{2}\left(1+\sqrt{\frac{M+2}{3M}}\right).
\label{FortM}
\vspace*{-2mm}
\end{equation}
We are now able to compare this fidelity to that of the optimal 
cloner for a pair of identical qubits $F_{||}(M)$:
for $M<6$, we have $F_{||}(M)>F_{\perp}(M)$, 
while $F_{||}(6)=F_{\perp}(6)$ and $F_{\perp}(M)>F_{||}(M)$ for $M>6$.
Thus, the cloner (\ref{cloner}) outperforms the standard cloner
for $M>6$. We note also that for $M\rightarrow \infty$,
the fidelity $F_{\perp}(M)$ tends to the optimal measurement 
fidelity $F_\perp$, as expected.

The optimality of the cloner can be proved with the help of techniques
adapted from the theory of semidefinite programming \cite{Audenaert01}.
We observe that the trace of
Lagrange multiplier $\lambda$ provides an upper bound on the achievable
fidelity. If $\lambda\otimes \openone_{\cal{K}}-A\geq 0$
then it holds for any $\chi$ that
${\rm Tr}_{{\cal{H}},\cal{K}}
[\chi \lambda\otimes \openone_{\cal{K}}]
\geq {\rm Tr}_{{\cal{H}},\cal{K}}[\chi A]$.
It follows from the unitarity constraint ${\rm
Tr}_{\cal{K}}[\chi]=\openone_{\cal{H}}$  that
${\rm Tr}_{{\cal{H}},\cal{K}} [\chi \lambda\otimes \openone_{\cal{K}}]={\rm
Tr}_{\cal{H}}[\lambda]$ does not depend on $\chi$.  Thus it holds that
${\rm Tr}_{\cal{H}}[\lambda]\geq {\rm Tr}_{{\cal{H}},\cal{K}}[\chi A]$.
From the numerical solution of Eqs. (\ref{extremaleq}) we have in basis
$|00\rangle$, $|11\rangle$, $|01\rangle$, $|10\rangle$,
\begin{equation}
\lambda=\frac{F_\perp(M)}{6}\left(
\begin{array}{cccc}
1 & 0 & 0 & 0 \\
0 & 1 & 0 & 0 \\
0 & 0 & 2 & -1 \\
0 & 0 & -1 & 2
\end{array}
\right).
\end{equation}
The block-diagonal matrix $\lambda\otimes \openone_{\cal{K}}-A$ is positive
semidefinite and has three different eigenvalues which read
$\mu_1=\frac{1}{12}\sqrt{\frac{M+2}{3M}}$,
$\mu_2=\frac{1}{3}\sqrt{\frac{M+2}{3M}}$, and
$\mu_3=0$. Since the upper bound ${\rm Tr}_{\cal{H}} [\lambda]=F_\perp(M)$
is saturated by our cloning machine, we conclude that our cloner is optimal.

In the rest of this paper, we will show that the cloning transformation
(\ref{cloner}) can be implemented probabilistically via stimulated parametric
down-conversion. The experimental setup under consideration
is shown in Fig. 1. This scheme is a straightforward  extension of the
setup suggested by Simon {\em et al.} \cite{Simon00} where
the qubits are represented by the polarization state of photons. 
We can identify $|0\rangle$ with vertical
polarization and $|1\rangle$ with horizontal polarization states.
In optical parametric down-conversion, a `blue' photon can split into
a pair of `red' photons. Traditionally, these daughter photons are
referred to as signal and idler, respectively. In our setup, three
nonlinear crystals $C_1$, $C_2$, $C_3$ are pumped by a strong laser beam.
In crystals $C_1$ and $C_2$, pairs of photons can be produced, so 
we can verify the presence of signal photons by detecting the idler photons 
emerging from $C_1$ and $C_2$.  If a single idler photon 
is detected on each side, then we have one signal photon in each beam.
The states of these two photons can
be manipulated with the help of phase shifters and polarization rotators
in order to prepare the desired input state $|\psi,\psi_\perp\rangle$.
The two photons then feed the signal and idler modes
of a third nonlinear crystal $C_3$, where $M$ clones are generated due to
the stimulated parametric down-conversion.

\begin{figure}[bh]
\centerline{\psfig{figure=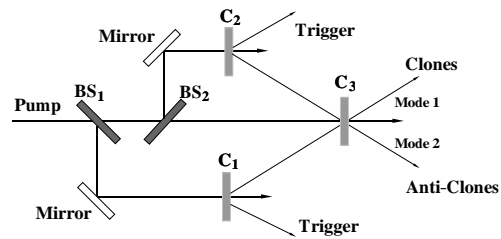,width=0.75\linewidth}}
\vspace*{2mm}
\caption{Setup for the cloning of orthogonal qubits via stimulated
parametric down-conversion. For a detailed description, see text.}
\end{figure}

In the limit of strong coherent pumping, the effective Hamiltonian describing
the interaction in $C_3$ can be written as follows \cite{Simon00},
\begin{equation}
H=i\hbar g(a_{V1}^\dagger a_{H2}^\dagger - a_{H1}^\dagger a_{V2}^\dagger)
+{\rm h.c.},
\label{Hamiltonian}
\end{equation}
where $a_{V1}^\dagger$ and $a_{H1}^\dagger$ denote bosonic creation operators
for photons in the first mode with vertical (V) or horizontal (H) polarization,
and similarly $a_{V2}^\dagger$ and $a_{H2}^\dagger$ are creation
operators for photons in the second spatial mode. The constant $g$ denotes
the parametric gain. The time evolution 
is thus governed by the unitary transformation $U=\exp(-iHt/\hbar)$.
With the help of the disentangling theorem, 
we can write the operator $U$ in a factorized form
\begin{eqnarray*}
U&=& e^{\Gamma a_{V1}^\dagger a_{H2}^\dagger} (\cosh
\gamma)^{-(a_{V1}^\dagger a_{V1}+a_{H2}^\dagger a_{H2}+1)}
e^{-\Gamma a_{V1} a_{H2}}
\nonumber \\
&&\times
e^{-\Gamma a_{H1}^\dagger a_{V2}^\dagger} (\cosh
\gamma)^{-(a_{H1}^\dagger a_{H1}+a_{V2}^\dagger a_{V2}+1)}
e^{\Gamma a_{H1} a_{V2}},
\nonumber
\end{eqnarray*}
where $\gamma=gt$ and $\Gamma=\tanh \gamma$.
The Hamiltonian (\ref{Hamiltonian}) has the important property of being
invariant under general simultaneous SU(2) transformations on the
polarization vectors ($a_V$, $a_{H}$) for modes $1$ and $2$ \cite{Simon00}.
It is thus sufficient to consider the evolution of a basis state
$|1\rangle_{V1}|0\rangle_{H1}|0\rangle_{V2}|1\rangle_{H2}$
(a single vertically polarized photon in mode $1$ and a single
horizontally polarized photon in mode $2$) which
represents the input state $|\psi,\psi_\perp\rangle \equiv |01\rangle$.
Making use of the factorized form of $U$,
we obtain the state at the output of the crystal $C_3$ in the form
\begin{eqnarray}
&&\sum_{M=0}^\infty \Gamma^{M-1}(1-\Gamma^2)
 \sum_{j=0}^M (-1)^{j} \left[(M-j)(1-\Gamma^2)-\Gamma^2\right]
 \nonumber  \\
&&\times |M-j\rangle_{V1}\, |j\rangle_{H1}\, |j\rangle_{V2}\, |M-j\rangle_{H2},
\label{output}
\end{eqnarray}
where $|k\rangle_l$ with $l=V1,H1,V2,H2$ denote the usual Fock states.
For a fixed number $M$ of photons in each mode $1$ and $2$,
the output state (\ref{output})
closely resembles the output state of the universal
cloning machine (\ref{cloner}) with the coefficients
$\alpha_{j,M}(\Gamma)\approx \left[(M-j)(1-\Gamma^2)-\Gamma^2\right](-1)^j.$
If we measure the number of photons in mode $2$ and detect $M$ photons,
then we know that $M$ photons representing $M$ approximate clones
of the input qubit $|\psi\rangle$ are present in mode $1$.
In order to calculate the fidelity of these clones, we insert
the properly normalized $\alpha_{j,M}(\Gamma)$  into formula (\ref{FortMsum}).
After some algebra, we obtain
\vspace*{-2mm}
\begin{equation}
\vspace*{-2mm}
F(M,y)=\frac{3 y^2-2y(2M+1)+\frac{3}{2}M(M+1)}%
{6y^2-6My+M(2M+1)}
\label{FidelDCy}
\end{equation}
where we have introduced $y=\Gamma^2/(1-\Gamma^2)\equiv \sinh^2\gamma$ 
for notational convenience. The cloning fidelity thus depends on the
parametric gain $\gamma$, so we must optimize this gain in order to achieve
the highest possible fidelity. Upon solving
$\frac{\partial F(M,y)}{\partial y}=0$ for $y$, we find that
\begin{equation}
y_{\rm opt}=\frac{M}{2}-\frac{1}{2}\sqrt{\frac{M(M+2)}{3}}.
\label{yopt}
\end{equation}
By inserting $y_{\rm opt}$ into Eq. (\ref{FidelDCy}),
we recover the optimal fidelity (\ref{FortM}).
Furthermore, it can be verified by direct calculation 
that with the optimal gain,
the postselected $M$-photon state at the output of the crystal
$C_3$ coincides with the output of the cloning machine (\ref{cloner}).

This approach of cloning based on down-conversion 
can be further extended 
to the approximate realization of the general cloning transformation
$\ket{\psi}^{\otimes N}\ket{\psi_{\perp}}^{\otimes N'} 
\to \ket{\psi}^{\otimes M}$. For $N'=1$, we have been able
to derive the optimal fidelity 
for any $N$ and $M\geq N$ 
by a similar calculation,
\begin{equation}
F_{\perp}(N,M)={N+1 \over N+3}+{3(N-1)+\sqrt{P/(N+2)} \over 2M(N+3)}
\end{equation}
with $P=(N-1)(N^2-15N-18)+8M(N+1)(M+3-N)$.
It can be checked that there is again a value of $M$ above which
this cloner outperforms the standard $(N+1)\to M$ cloner. For large
$N$, however, the advantage becomes marginal.

In summary, we have designed a universal cloning machine for
orthogonal qubit pairs, and have shown that it achieves
a higher fidelity for $M>6$ than the standard cloner 
for identical qubits. 
We conclude that
the advantage of orthogonal qubits over identical qubits
that was discovered in the context of measurement
also extends to cloning.

N.J.C. is grateful to Christoph Simon for helpful discussions.
J.F. acknowledges support from the grant No LN00A015 of the Czech Ministry 
of Education. S.I. is supported by a fellowship from the Belgian 
FRIA foundation. N.C and S.M. acknowledge funding by the European Union
under project IST-FET-EQUIP.

\end{document}